\documentclass[twocolumn]{aastex63}

\usepackage{lineno}
\usepackage[figuresright]{rotating}

\received{***, ****}
\revised{***, ****}
\accepted{***, ****}
\submitjournal{ApJ}

\shorttitle{Toward a Unified Explanation}
\shortauthors{Song et al.}

\begin{document}
\title{Toward a Unified Explanation for the Three-part Structure of Solar Coronal Mass Ejections}

\correspondingauthor{Hongqiang Song}
\email{hqsong@sdu.edu.cn}

\author{Hongqiang Song}
\affiliation{Shandong Provincial Key Laboratory of Optical Astronomy and Solar-Terrestrial Environment, and Institute of Space Sciences, Shandong University, Weihai, Shandong 264209, China}
\affiliation{CAS Key Laboratory of Solar Activity, National Astronomical Observatories, Chinese Academy of Sciences, Beijing, 100101, China}








\author{Leping Li}
\affiliation{CAS Key Laboratory of Solar Activity, National Astronomical Observatories, Chinese Academy of Sciences, Beijing, 100101, China}

\author{Yao Chen}
\affiliation{Institute of Frontier and Interdisciplinary Science, Shandong University, Qingdao, Shandong 266237, China}
\affiliation{Shandong Provincial Key Laboratory of Optical Astronomy and Solar-Terrestrial Environment, and Institute of Space Sciences, Shandong University, Weihai, Shandong 264209, China}








\begin{abstract}
Coronal mass ejections (CMEs) are associated with the eruption of magnetic flux ropes (MFRs), which usually appear as hot channels in active regions and coronal cavities in quiet-Sun regions. CMEs often exhibit the classical three-part structure in the lower corona when imaged with white-light coronagraphs, including the bright front, dark cavity, and bright core. The bright core and dark cavity have been regarded as the erupted prominence and MFR, respectively, for several decades. However, recent studies clearly demonstrated that both the prominence and hot-channel MFR can be observed as the CME core. The current research presents a three-part CME resulted from the eruption of a coronal prominence cavity on 2010 October 7 with observations from two vantage perspectives, \textit{i.e.,} edge-on from the Earth and face-on from the \textit{Solar Terrestrial Relations Observatory (STEREO)}. Our observations illustrates two important results: (1) For the first time, the erupting coronal cavity is recorded as a channel-like structure in the extreme-ultraviolet passband, analogous to the hot-channel morphology, and is dubbed as warm channel; (2) Both the prominence and warm-channel MFR (coronal cavity) in the extreme-ultraviolet passbands evolve into the CME core in the white-light coronagraphs of \textit{STEREO-A}. The results support that we are walking toward a unified explanation for the three-part structure of CMEs, in which both prominences and MFRs (hot or warm channels) are responsible for the bright core.
\end{abstract}



\section{Introduction}
Coronal mass ejections (CMEs) are an energetic eruption occurred in the solar atmosphere \citep{forbes00,chenpengfei11,webb12,chengxin17,guoyang17} and can lead to disastrous space weather \citep{gosling91,zhangjie03,zhangjie07,xumengjiao19}. Since the first observation of CMEs with the space-borne coronagraph on 1971 December 14 \citep{tousey73}, a lot of progress has been achieved on various aspects of CMEs through remote-sensing and in-situ observations \citep{song20b}, including the precursor, energy source, trigger mechanism, morphology, velocity, mass, occurrence rate, propagation process, as well as their influences on the earth space environment and human high-technology activities \citep{chenpengfei11,webb12}. While we do not have the final conclusions yet for many issues of CMEs due to their complexity and our limited observations, including their morphological structure \citep{howard17}.

CMEs often exhibit a typical three-part structure in the white-light coronagraphs, \textit{i.e.}, the bright front, dark cavity, and bright core \citep{illing85, webb87,cremades04,vourlidas13}. A recent study demonstrated that CMEs, with and without the three-part structure in the white-light images, can possess the three-part appearance in the extreme-ultraviolet (EUV) passbands \citep{song19b}. This indicates that the three-part structure might be an intrinsic feature of CMEs in their early eruption stage. 

The structure of CMEs should be tightly correlated with their origins. Theoretical studies \citep[e.g.,][]{mikic94,gibson98,amari00} propose that CMEs originate from the eruption of magnetic flux ropes (MFRs), which are a coherent magnetic structure with magnetic field lines twisting more than 1 turn and can form both prior to \citep{tripathi09,green09,chengxin11,zhangjie12,patsourakos13,song15b,kliem21} and during \citep{song14a,ouyang15,wangwensi17,jiangchaowei21} eruptions. Usually, MFRs appear as hot channels with temperature beyond 10 MK when erupting from active regions \citep{chengxin11,zhangjie12,tripathi13,song14b,song15c,nindos15,aparna16}, while in quiet-Sun regions, they appear as coronal cavities that correspond to the cross-section of large scale quiet Sun MFRs or filament channels \citep{marque04,wangyimin08,bak-steslicka13,karna15b,chenyajie18}.

The magnetic dips of MFRs can support prominences (called filaments when located on the solar disk) against gravity and maintain them suspending in the corona. Prominences are cooler and denser than the coronal plasma by about two orders of magnitude \citep{parenti14,yanxiaoli16,song17a,wangjiemin18}. MFRs can erupt taking away prominences together, so more than 70\% of CMEs are associated with prominence eruptions \citep{webb87,gopalswamy03}. This enables us to study the dynamic process of MFRs through prominences \citep{qiujiong04,song13,song15a,song18a,song18b,chengxin20}.

As the K-corona originates from the Thomson scattering of free electrons, its brightness depends on the electron column density \citep{hayes01}. Therefore, the three-part appearance of CMEs in coronagraphs implies that they possess a high-low-high density structure, which has been explained as the observational manifestations of background plasma pileup (high density), MFR (low density), and prominence (high density) for several decades.

However, recent studies challenged the above traditional explanation \citep{howard17,song17b,song19a,song19b}. \cite{howard17} conducted a statistics on 42 three-part CMEs and found that 69\% of them are unrelated with filament eruptions. They speculated that the bright core results from the geometric projection of MFRs. Multi-perspective observations demonstrated that both filaments and hot-channel MFRs can evolve into the core \citep{song17b,song19b}, and in some situations, one CME can possess both a sharp and a fuzzy core, corresponding to the filament and MFR, respectively \citep{song19a}. Therefore, \cite{song19a} proposed a new explanation for the three-part structure of CMEs, which suggests that both prominences and MFRs can be recorded as the bright core in the early eruption stage, and the dark cavity corresponds to a low-density zone with sheared magnetic fields between the leading front and MFR, see Figure 6 in \cite{song19a}.

The above explanation has been confirmed well for hot-channel MFRs no matter whether prominences are involved or not \citep{song17b,song19b}. Then a subsequent question arises: does the new scenario hold for the CMEs related to coronal-cavity MFRs? Coronal cavities typically exist below 1.6 $R_\odot$ \citep{maricic04,gibson06b}, and contain flows with speeds of 5--10 km s$^{-1}$ and scales of tens of megameters \citep{schmit09}. The density of prominences is obviously higher than their surroundings as mentioned, while the cavities have a density depletion compared to their rim and overlying streamer \citep{marque04,fuller09}. Can the coronal cavities evolve into the bright core of CMEs, \textit{i.e.}, can we unify the explanation for the three-part structure of CMEs wherever they originate from active regions or quiet-Sun regions?

In this paper, we address this question by means of observations from two orthogonal perspectives, which demonstrate that the new scenario can apply for CMEs resulted from coronal cavity eruptions. The paper is organized as follows. In Section 2 we introduce the related instruments, and the observational results are displayed in Section 3. Section 4 presents the discussion, which is followed by the summary in the final section.

\section{Instruments}
The event is analyzed with multiple instruments, including the Atmospheric Imaging Assembly \citep[AIA;][]{lemen12} on board the Solar Dynamics Observatory \citep[SDO;][]{pesnell12}, the Large Angle and Spectrometric Coronagraph \citep[LASCO;][]{brueckner95} on board the Solar and Heliospheric Observatory \citep[SOHO;][]{domingo95}, as well as the Extreme Ultraviolet Imager (EUVI) and white-light coronagraphs (COR1 and COR2) of Sun Earth Connection Coronal and Heliospheric Investigation \citep[SECCHI;][]{howard08} on board the Solar Terrestrial Relations Observatory \citep[STEREO;][]{kaiser05}.

SDO is located in a non-equatorial geosynchronous orbit around the Earth, which has an inclination of $\sim$29$^{\circ}$ and an altitude of $\sim$35,000 km. SOHO is in an elliptical Lissajous orbit around the L1 libration point, which has a distance of $\sim$1.5 million km from the Earth. Both SDO and SOHO observe the Sun from the Earth perspective. STEREO consists of twin spacecraft orbiting the Sun, with one ahead (A) of and the other behind (B) the Earth. The orbital periods of STEREO-A and B are slightly shorter and longer than the Earth's, respectively. This makes the twin spacecraft separate from the Earth in opposite directions, thus STEREO can observe the Sun from different perspectives and allow us to analyze the three-dimensional eruption of coronal prominence cavities.

The field of view (FOV) of AIA is 1.3 $R_\odot$, and it can image the corona in seven EUV passbands with high temporal cadence (12 s) and spatial resolution (1.2{\arcsec}). The 193 \AA\ and 304 \AA\ passbands are used to analyze the coronal cavity and prominence, respectively. The LASCO includes three coronagraphs and covers a large FOV ranging from 1.1 to 30 $R_\odot$, while only C2 (2.2--6 $R_\odot$) and C3 (4--30 $R_\odot$) are available since 1998. This leaves a FOV gap between the AIA and LASCO. The FOV of EUVI is 1.7 $R_\odot$, partially overlapping with that of COR1 (1.4--4 $R_\odot$), which enables us to observe the erupted prominence seamlessly from solar surface to the outer corona, although with a relatively low cadence (3--5 minutes for 195 \AA\ and 10 minutes for 304 \AA) and resolution (2.4{\arcsec}).

\section{Observations}
An eruption of coronal prominence cavity and the associated three-part CME were recorded by SDO, SOHO, and STEREO-A on 2010 October 7 when STEREO-A was $\sim$83$^{\circ}$ west of the Earth as displayed in Figure 1. It is a limb event from the Earth perspective while a disk one for STEREO-A. The nearly orthogonal perspectives provide us an excellent opportunity to analyze the nature of the three-part structure of CMEs related to the coronal cavity.

Figure 2 displays the observations from the Earth direction. Panel (a) shows the prominence recorded by AIA 304 \AA\ prior to eruption at 03:19:34 UT. The white line depicts its projection shape. Panel (b) presents the coronal cavity through AIA 193 \AA\ at 03:19:32 UT, which is reduced with the radial filter code of solar software $aia\_rfilter.pro$. The code enhances the coronal contrast by summing a number of solar images and dividing the coronal component into rings. Then each ring is scaled based on its radius and pixel brightness relative to its neighbors. Note that the disk is pulled from the central image to create the final output. The red dots delineate the outer rim of the cavity. The top rim is outside of the AIA FOV, which prevents us from tracing the cavity motion quantitatively in the inner corona. The coronal cavity indicates edge-on observations from the SDO and SOHO, \textit{i.e.,} the line of sight is parallel to the MFR axis. The prominence shape in Panel (a) is superimposed on Panel (b) to demonstrate their location relation. The prominence approximately locates at the cavity bottom, while seems a little lower because the cavity tube is curvy circling the solar surface \citep{karna15a} and the prominence does not just suspend above the limb.

The cavity begins to ascend slowly from $\sim$07:00:00 UT onward, taking away the prominence together. See the accompanying animation to inspect the complete eruption process. The eruption of coronal cavity leads to a three-part CME that is observed by the LASCO as shown in Panel (c), where the bright front is delineated with blue dots, the dark cavity and bright core are denoted with the leftward and upward arrows, respectively. The CME first appears in the C2 FOV at 7:24:05 UT and its linear speed is 417 km s$^{-1}$ (CDAW\footnote{https://cdaw.gsfc.nasa.gov}). In the inner corona, the post-eruptive arcades \citep[PEA;][]{tripathi04} imaged with the AIA 193 \AA\ passband is obvious as pointed by the white arrow in Panel (d).

It is difficult to examine the correspondence between the coronal-cavity MFR and the CME core by means of the AIA and LASCO due to the FOV gap between them, thus we turn to the face-on observations of the STEREO-A, \textit{i.e.,} the line of sight is perpendicular to the MFR axis, as shown in Figure 3. Panel (a) presents the erupting prominence imaged by the EUVI 304 \AA\ at 09:46:15 UT with the red line depicting its projection shape. Note that this prominence does not exhibit observable filament feature on the disk of EUVI 304 \AA\ prior to eruption, possibly due to its small scale and/or high altitude. The prominence signal becomes weak gradually when propagating outward, thus a base-difference image of 304 \AA\ at 11:16:15 UT is employed to demonstrate its motion and position as shown in Panel (b), in which Panel (a) is subtracted as the base image. The red line delineates the prominence position and shape at this time.

The CME also possesses a three-part structure in the FOVs of both COR1 and COR2 as shown in Panels (c) and (d). Panel (c) illustrates the base-difference image of COR1 observation at 11:15:09 UT (with 06:05:09 UT as the base time). The CME front is relatively weak in COR1 images and is described with the blue dots. The prominence shape (red line) in Panel (b) is over-plotted on COR1 image as pointed with the white arrow. This prominence can be tracked continuously to COR2 FOV, and corresponds to the sharp and brighter core portion as delineated with green line and pointed with the arrow in Panel (d). See the accompanying animation (left part in each frame) to inspect the correspondence. The length (width) of the prominence increases from $\sim$343{\arcsec} (86{\arcsec}) at 11:15:09 UT (Panel (c)) to $\sim$1190{\arcsec} (543{\arcsec}) at 13:54:00 UT (Panel (d)), which demonstrates the prominence expansion during its propagation. Panel (d) presents the three-part CME at 13:54:00 UT, in which both the leading front and the core are clear. The white-light images of both COR1 and COR2 illustrate that the area of the prominence projection only occupies a small portion of the CME core. Therefore, the prominence can not be responsible for the whole core, or there still exists a core when the prominence is not involved, consistent with a series of previous observations \citep{howard17,song17b,song19a,song19b}.

As mentioned above, the new explanation for the three-part CMEs suggested that MFRs can be observed as the bright core, which has been confirmed by the hot-channel MFRs \citep[e.g.,][]{song19b}. To connect the coronal-cavity MFR with the CME core, we need to find the erupted MFR recorded by the STEREO-A. No filament channel can be observed in the EUVI 195 \AA\ image prior to eruption, while an ascending diffuse channel structure (delineated with the red dots) appears in the base-difference image of EUVI 195 \AA\ during eruption as shown in Panel (e). This structure is almost not discernible in the static image due to its weak emission, while the accompanying animation (right part in each frame) can illustrate its existence and moving outward. The prominence projection shape in Panel (a) is over-plotted in Panel (e) as shown with the red line, demonstrating that the prominence locates at the bottom of the channel, consistent with the relation between prominence and coronal-cavity MFR in Figure 2(b). In the meanwhile, the channel width is $\sim$300{\arcsec}, close to the projection width of the coronal cavity when viewed from STEREO-A perspective. All these suggest that the EUVI channel correspond to the face-on observation of the AIA coronal cavity, \textit{i.e.,} the channel is the MFR. The composite observations of EUVI, COR1 and COR2 demonstrate that the channel evolved into the fuzzy core of the CME in COR1 and COR2 images, also see the accompanying animation (left part in each frame). After the channel eruption, the PEAs can be observed by the EUVI 195 \AA\ as pointed with two arrows in Panel (f).

To the best of our knowledge, this should be the first time to observe an erupting coronal cavity as a coherent channel from quiet-Sun regions. The coronal cavities are filled with plasma with temperatures being a few MK, \textit{e.g.,}1.4--1.7 MK \citep{kucera12}, 1.67--2.15 MK \citep{bak-steslicka13}, which are not as high as the hot channel ($\sim$10 MK or beyond) erupted from active regions \citep[e.g.,][]{zhangjie12}, and not as low as the prominences (such as 4000 - 20000 K) suspending in the corona \citep{park13}. The coronal-cavity temperature is a little higher than the surrounding corona \citep{fuller08,vasquez09,habbal10,reeves12,karna15a}, thus the erupting channel from the quiet-Sun region is dubbed as warm channel.

Figure 4 displays the dual-viewpoint images to further illustrate the correspondence of prominence, MFR, and three-part CME between the observations of SDO or SOHO and STEREO-A. These images are created with JHelioviewer\footnote{http://www.jhelioviewer.org/} that is a visualization software for solar image data based on the JPEG 2000 compression standard \citep{muller17}. Panel (a) presents the composite image of AIA 304 and 193 \AA\ from the Earth direction, as well as the composite image of EUVI 304 and 195 (base-difference) \AA\ around 09:27 UT. Note that the coalignment of the images from different instruments is conducted by the JHelioviewer. The upward and rightward arrows depict the prominence recorded by the AIA and EUVI, respectively. As shown above, the coronal cavity is obvious in the static image of AIA 193 \AA, while almost unidentifiable when imaged face-on with the EUVI 195 \AA. The accompanying animation (left part in each frame) displays the eruption process from two perspectives simultaneously, and demonstrate that the coronal cavity and warm channel move outward synchronously, further confirming that they are the different projections of the tubular MFR \citep{karna15a}.

Panel (b) of Figure 4 presents the dual-viewpoint image with a larger FOV, adding the white-light coronagraph LASCO (C2 and C3), as well as COR1 and COR2. This panel demonstrates straightforward correspondences of the three-part structure observed from different perspectives, especially for the bright front and dark cavity. See the accompanying animation (right part in each frame) to examine the correspondences continuously. The CME core in the LASCO image (as depicted with the upward arrow) is bright and obviously larger than the sharp core (the prominence, as depicted with the rightward arrow) in the COR2 image. For edge-on observation of MFR, the MFR axis is parallel to the line of sight, along which the projections of the MFR and streamer make the prominence portion unidentifiable from the entire core.

Figure 5 elaborates our observation and explanation on this three-part CME through a cartoon, where the bright front and MFR (core) are described with blue and green, respectively, and the red represents the prominence (denser core). In this event, the prominence length is much shorter compared to the MFR. Panel (a) is the edge-on observation from the SDO and SOHO perspective. As mentioned, the coronal cavity (\textit{i.e.}, the cross section of the large scale quiet Sun MFR) has a density depletion compared to its rim, which is expressed with different color saturations. The mean depletion is 28\% at the 1.2 $R_\odot$ according to a survey based on 24 coronal cavities \citep{fuller09}. While coronal cavities have higher density than their both sides along the solar limb, such as the coronal holes and equatorial regions \citep{koutchmy92,fuller09}. This makes that the coronal-cavity MFR can be imaged as the bright core of CMEs due to the existence of a low-density zone between the MFR and the front, see discussion in Section 4. Panel (b) represents the face-on observation of the STEREO-A. The MFR appears as a coherent channel and is responsible for the large CME core, containing a small and brighter core (the high-density prominence).

\section{Discussion}
When the MFR ascends from its source region, it expands and stretches its overlying loops successively, and the background plasmas are compressed or piled up along the loops, which evolve into the bright front \citep{forbes00,chenpengfei09}. Numerical simulations illustrates that the MFR can not fill the full space below the loops during the onset \citep{antiochos99,lynch08,fanyuhong16}, supported by the observations that show a low-density zone existing between the hot-channel MFR and overlying loops \citep[e.g.,][]{song14b}. \cite{haw18} suggested that a rising electric current in the MFR can induce a directed electric current with opposite direction in the surrounding corona, then the magnetic force between the inner-MFR electric current and the induced current propels the coronal plasma away from the MFR, leading to the formation of the low-density zone. This zone will be imaged as the dark cavity of three-part CME due to its relatively low density. Our current observations demonstrate that the warm-channel MFR (coronal cavity) can also appear as the CME core like the hot-channel MFR \citep{song19b}. Therefore, a unified explanation for the three-part structure of CMEs might be acquired in spite of their originating from active or quiet-Sun regions, which suggests that MFRs can evolve into the CME core, in addition to prominences.

We point out that the prominence and coronal-cavity MFR could be regarded as the bright core and dark cavity of three-part CMEs in three cases at least. First, the density of the CME leading front is relatively low, thus the front is weak and unidentifiable for the low-sensitivity coronagraphs; Second, the coronagraphic FOV is small and only the square region is imaged as delineated with the dotted lines in Figure 5(a). In these two cases, the coronal cavity rim might be observed as the bright front with white-light coronagraphs; Third, the low-density zone is occupied by the expanding and growing coronal-cavity MFR, see Figure 6(c) of \cite{song19a}. In this case, the coronal-cavity rim merges with the leading front.

Just like the erupting hot channel can originate from the sigmoid prior to eruption \citep{liurui10}, the warm channel during eruption might correspond to the filament channel before eruption. No filament channel is observed for this event, possibly because the altitude of the coronal cavity is relatively high as shown in Figure 2(b).

\section{Summary}
In this paper, we report a three-part CME occurred in a quiet-Sun region on 2010 October 7, which is induced by a coronal prominence cavity eruption. This event was recorded by AIA and LASCO from the Earth perspective, as well as EUVI and COR from the STEREO-A perspective. The dual-viewpoint observations provide us an excellent opportunity to analyze the nature of the three-part structure as both perspectives are almost orthogonal with a separation angle of $\sim$83$^{\circ}$.

It is a limb event viewed from the Earth direction, and the AIA 193 \AA\ imaged the coronal cavity and its eruption process due to the edge-on view. For the first time, the erupting coronal cavity is observed as a warm channel from the face-on view in the EUVI 195 \AA\ passband. Thanks to the seamlessly observations provided by the EUVI, COR1, and COR2, the prominence and the warm channel can be traced continuously, which demonstrated that both the prominence and the channel (coronal-cavity MFR) evolved into the CME core.

Combined with our previous studies that illustrate the correspondence between the hot-channel MFR and the bright core \citep{song17b,song19b}, it is likely that we are walking toward a unified explanation for the three-part structure of CMEs, which claims that both prominences and MFRs can correspond to the bright core of CMEs in the early stage of eruptions. More efforts are needed to study the formations of the bright front and dark cavity of CMEs.

\acknowledgments We thank the referee for the comments and suggestions that helped to improve the original manuscript. Hongqiang Song thanks Profs. Jie Zhang (GMU), Xin Cheng (NJU), and Pengfei Chen (NJU) for their helpful discussions. This work is supported by the NSFC grants U2031109, 11790303 (11790300), and 12073042. Hongqiang Song is also supported by the CAS grants XDA-17040507 and the open research program of the CAS Key Laboratory of Solar Activity KLSA202107. The authors acknowledge the use of data from the \textit{SDO}, \textit {SOHO}, and \textit{STEREO} missions, as well as the usage of the JHelioviewer software.





\begin{figure*}[htb!]
\epsscale{0.8} \plotone{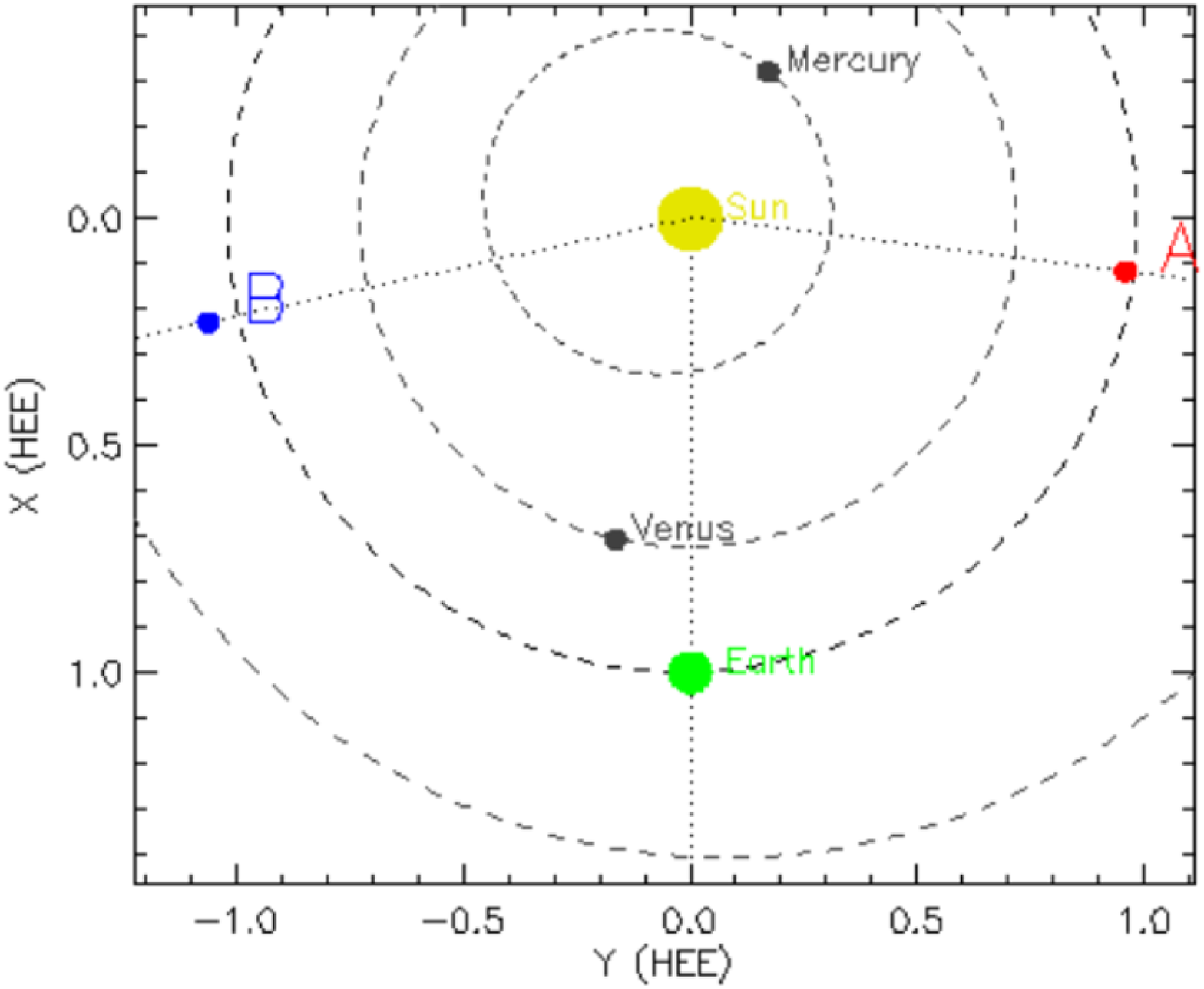} \caption{Positions of the Earth (SDO and SOHO) and STEREO in the ecliptic plane on 2010 October 7 (https://stereo.gsfc.nasa.gov/). STEREO-A is $\sim$83$^{\circ}$ west of the Earth. \label{Figure 1}}
\end{figure*}

\begin{figure*}[htb!]
\epsscale{0.9} \plotone{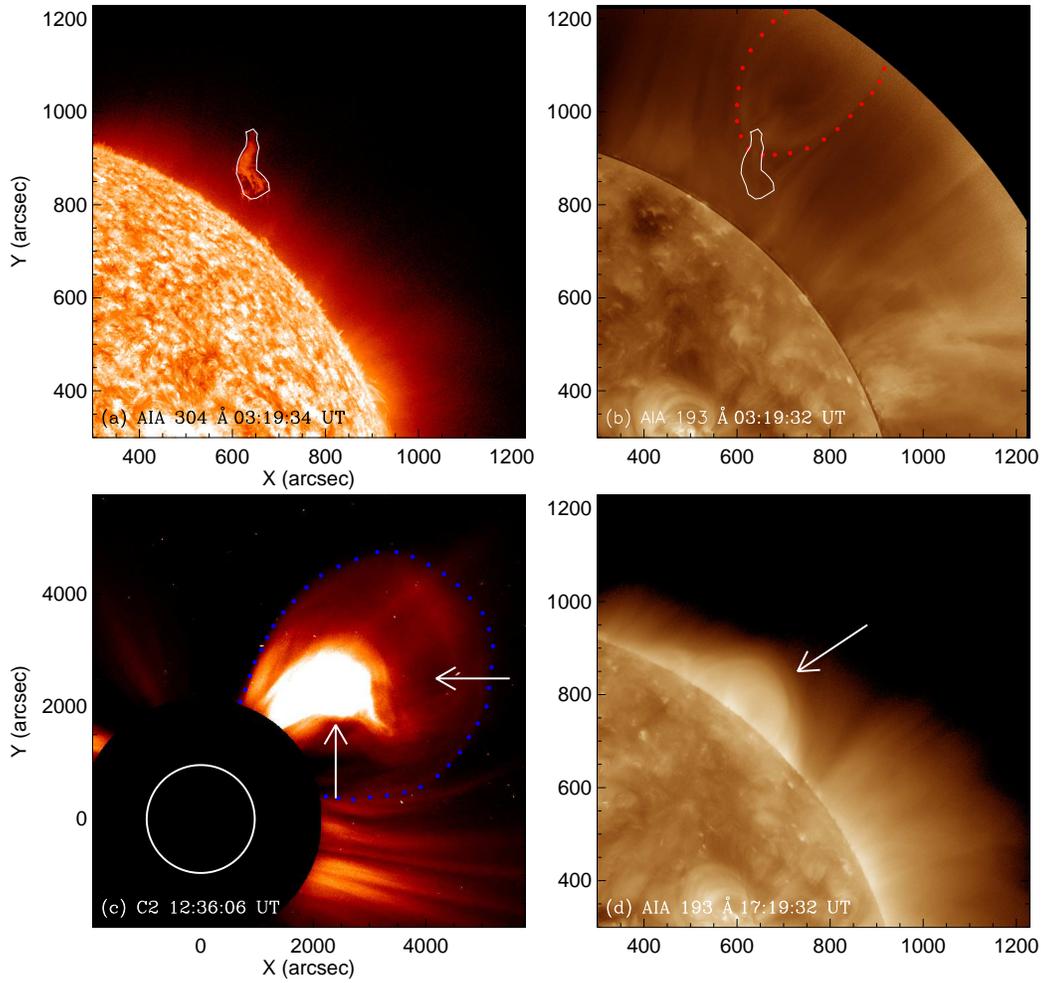} \caption{Edge-on observations from the Earth perspective. (a) AIA 304 \AA\ presents the prominence prior to eruption. (b) The enhanced AIA 193 \AA\ displays the coronal cavity with the prominence outline in Panel (a) over plotted. (c) The associated three-part CME recorded by LASCO C2. The front is delineated with blue dots, and the dark cavity and bright core are pointed with the leftward and upward arrows, respectively. (d) The PEA imaged by the AIA 193 \AA\ passband. Panels (a) and (b) are accompanied by an animation that displays the eruption process of the prominence and coronal cavity from 03:19 UT to 10:39 UT with each frame being a composite image of AIA 193 and 304 \AA. The real-time duration of the animation is 3 s. \\ (An animation of this figure is available.)\label{Figure 2}}
\end{figure*}

\begin{figure*}[htb!]
\epsscale{1.0} \plotone{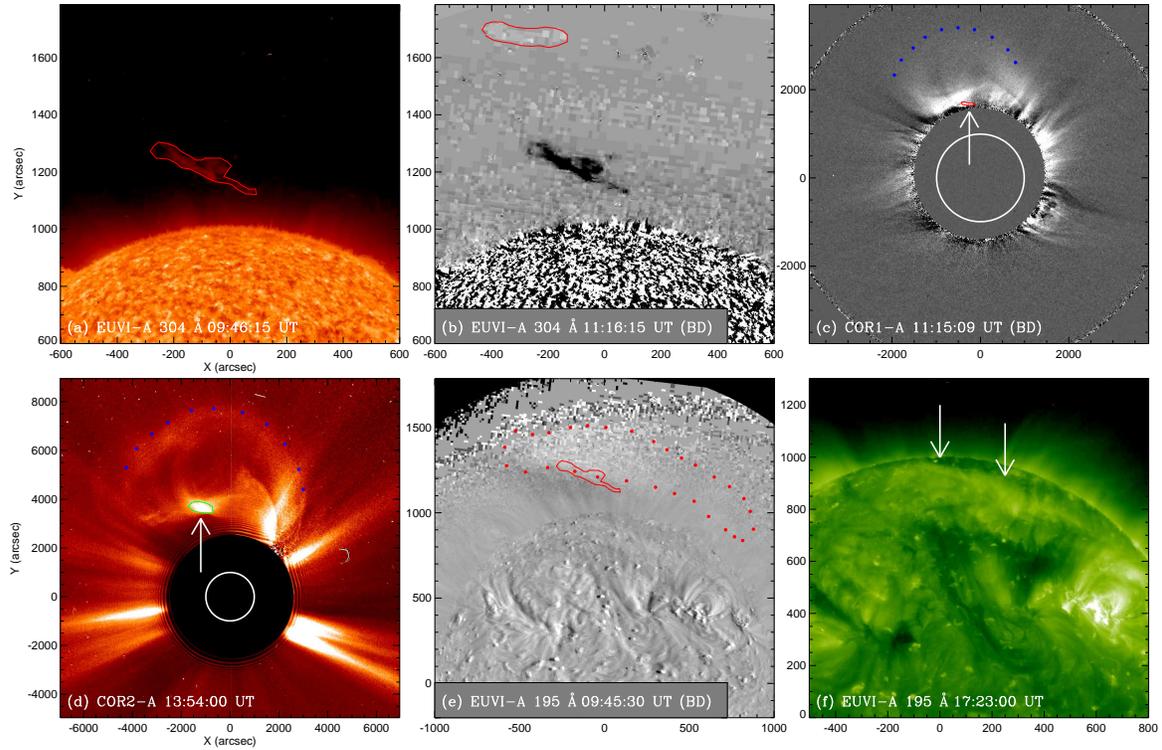} \caption{Face-on observations from the STEREO-A perspective. (a) EUVI 304 \AA\ shows the prominence during eruption. (b) Base-difference image of EUVI 304 \AA\ shows the new position of the erupting prominence. (c) Base-difference image of COR1 displays the three-part CME. (d) The CME in the COR2 image. (e) Base-difference image of EUVI 195 \AA\ shows the erupting MFR as the warm channel that is delineated with red dots. The prominence location in (a) is over-plotted with red line. (f) EUVI 195 \AA\ image presents the PEAs. The left part in each frame of the animation (accompanying Panels (a)--(e)) displays the MFR eruption taking away the prominence together from 07:00 UT to 15:00 UT with a composite observations of EUVI (304 and 195 \AA), COR1, and COR2, and the right part in each frame (accompanying Panel (e)) presents the moving process of the warm channel from 06:25 UT to 12:25 UT through base-difference images of EUVI 195 \AA. The duration of the animation is 4 s. \\ (An animation of this figure is available.)\label{Figure 3}}
\end{figure*}

\begin{figure*}[htb!]
\epsscale{1.0} \plotone{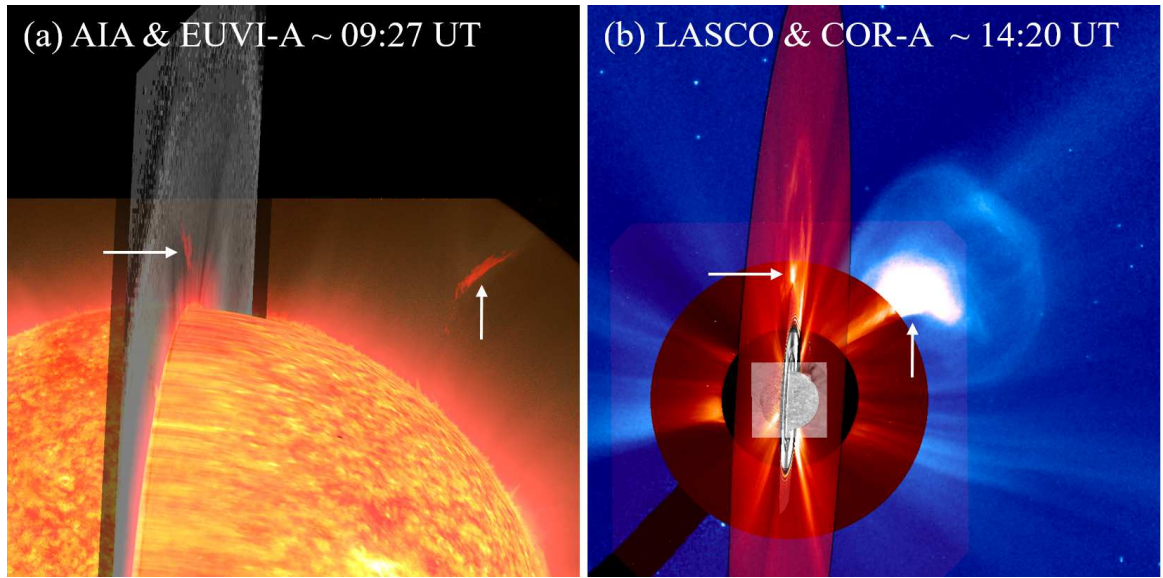} \caption{Dual-viewpoint images from the Earth and STEREO-A (observing from right) views. (a) The observation through the AIA (304 and 193 \AA) and EUVI (304 and 195 \AA), showing the prominence and coronal-cavity MFR. (b) The observation through the C2, C3, COR1, and COR2, showing the three-part CME. The left part in each frame of the animation (accompanying Panel (a)) displays the eruption process in the inner corona from 7:00 UT to 11:00 UT, and the right part in each frame (accompanying Panel (b)) displays the CME in the outer corona from 7:00 UT to 15:00 UT. The duration of the animation is 4 s. Both figures and animations are created with the JHelioviewer software \citep{muller17}. \\ (An animation of this figure is available.)\label{Figure 4}}
\end{figure*}

\begin{figure*}[htb!]
\epsscale{1.0} \plotone{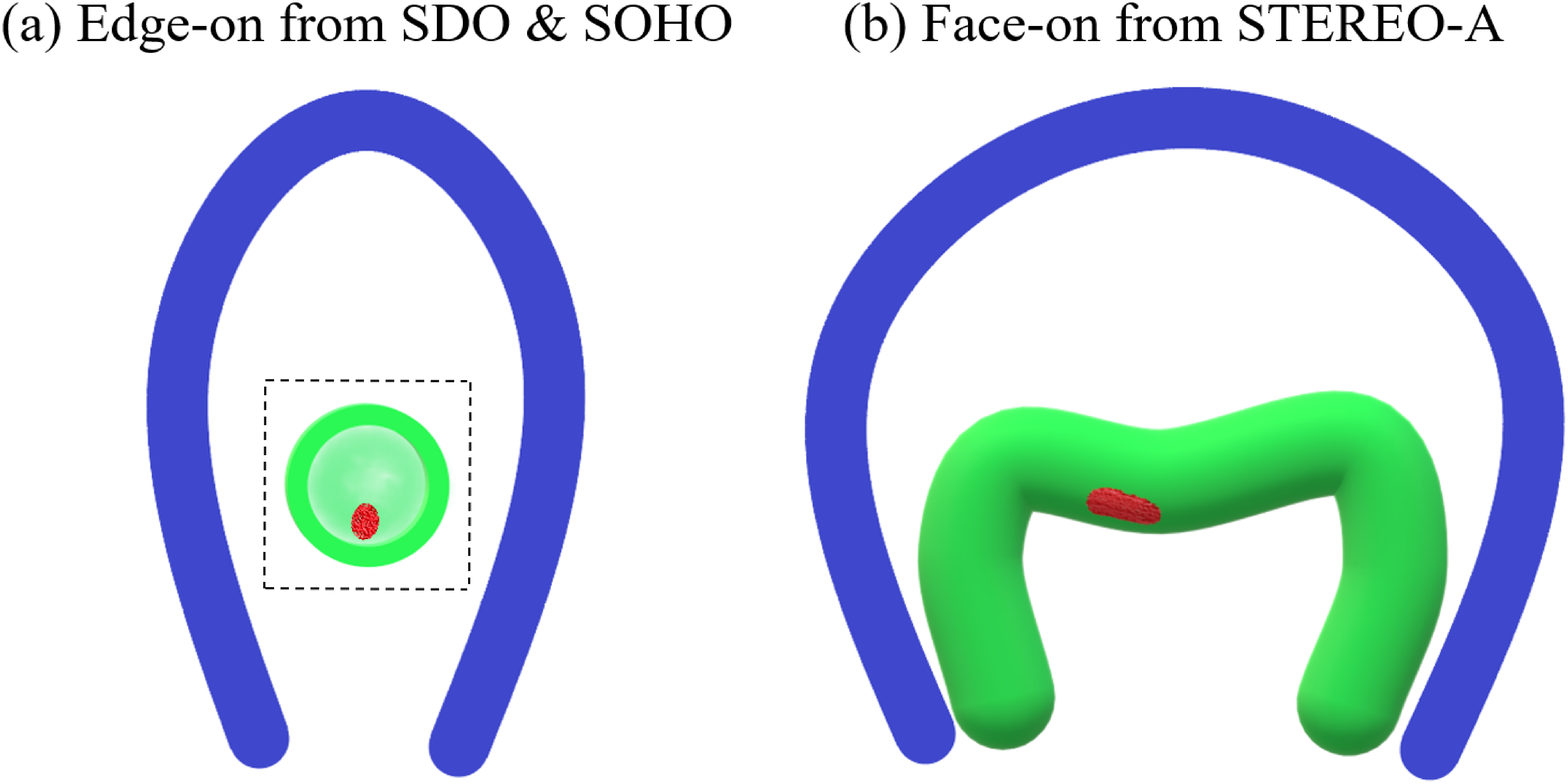} \caption{A schematic drawing of the unified explanation for three-part structure of CMEs. Panel (a) represents the observations from edge-on perspective, and Panel (b) for face-on. The coronal cavity or the warm-channel MFR (green) corresponds to the CME core, just like the hot-channel MFR. See text for details. \label{Figure 5}}
\end{figure*}

\end{document}